\documentclass[amsmath,amssymb,floatfix,lengthcheck,superscriptaddress]{revtex4-1}
\usepackage{graphicx}
\usepackage[centering,hmargin=20mm,tmargin=29mm,bmargin=25mm]{geometry}
\usepackage{multirow}
\usepackage{newtxtext}
\usepackage[cmintegrals]{newtxmath} 
\usepackage{mathtools}
\usepackage{amsmath,amssymb,amsfonts}
\usepackage{bm}
\usepackage{lineno}
\usepackage{soul}
\usepackage{siunitx}
\usepackage{xcolor}
\usepackage[pdfencoding=auto]{hyperref}
\usepackage{cleveref}
\hypersetup{colorlinks,
	linkcolor={blue!75!black!80!yellow},
	citecolor={blue!75!black!80!yellow},
	urlcolor={blue!75!black!80!yellow}
}

\begin{document}
\title{
Ultrafast altermagnetophononics
}
\author{Chenyu Wang}
\affiliation{MIIT Key Laboratory of Semiconductor Microstructure and Quantum Sensing Department of Applied Physics, Nanjing University of Science and Technology, Nanjing 210094, China}
\affiliation{Beijing National Laboratory for Condensed Matter Physics and Institute of Physics, Chinese Academy of Sciences, Beijing 100190, China}
\affiliation{School of Physical Sciences, University of Chinese Academy of Sciences, Beijing 100190, China}
\author{Yaxian Wang}
\email[Email address:\;]{yaxianw@iphy.ac.cn}
\affiliation{Beijing National Laboratory for Condensed Matter Physics and Institute of Physics, Chinese Academy of Sciences, Beijing 100190, China}
\author{Sheng Meng}
\email[Email address:\;]{smeng@iphy.ac.cn}
\affiliation{Beijing National Laboratory for Condensed Matter Physics and Institute of Physics, Chinese Academy of Sciences, Beijing 100190, China}
\affiliation{School of Physical Sciences, University of Chinese Academy of Sciences, Beijing 100190, China}
\affiliation{Songshan Lake Materials Laboratory, Dongguan, Guangdong 523808, China}

\begin{abstract}
Altermagnets feature symmetry-dictated nontrivial spin splitting in electronic structure promising for next-generation spintronics, yet their ultrafast dynamical manipulation remains largely unexplored.
Here, we establish altermagnetophononics as an efficient route for magnetic control via selective symmetry breaking induced by coherent phonons.
Using the prototypical altermagnet $\alpha$-MnTe as an example, we demonstrate that the targeted excitation of B$_{1g}$ mode selectively lifts the symmetry constraints protecting altermagnetism (AM), driving ultrafast transition into a transient compensated ferrimagnetic (cFiM) phase characterized by a global spin splitting without net magnetization. 
Further, we show that the proposed mechanism is broadly applicable by demonstrating a multi-mode symmetry breaking pathway, and by realizing a ferrimagnetic order with reversible magnetic moment in metallic CrSb.
These findings elucidate the potential to obtain desirable nonequilibrium properties in altermagnets via coherent phononic control over their spin splittings.

\end{abstract}
\date{\today}
\maketitle
Crystal symmetry plays a central role in determining the basic properties of quantum materials, dictating not only the existence of ordered phases but also the structure of their elementary excitations.
In magnetic systems, it governs the identification of altermagnet, characterized as a collinear antiferromagnet yet with momentum-dependent nonrelativistic spin splitting~\cite{AM202209,AM202212,AM2020RuO2,McClarty2024,Song2025}.
This unique symmetry-enforced spin texture gives rise to a rich array of time-reversal symmetry-breaking phenomena such as anomalous Hall transport, offering promising opportunities for spintronic applications that combine the robustness of antiferromagnets with functionalities typically associated with ferromagnets~\cite{AM2022AHE,Hariki2024,Gonzalez2021,Chen2024}.
Owing to the symmetry origin, altermagnets are inherently sensitive to structural modifications.
However, although the altermagnetic state can be effectively manipulated via static crystal symmetry engineering in thermal equilibrium~\cite{Zhou2025,Amin2024,Karetta2025,Yuan2024,Atasi2024,Wang2025}, discussions on extending the symmetry control strategies to ultrafast timescales are rather limited.
Identifying efficient pathways to dynamically modulate altermagnetism is therefore a key step toward realizing high-speed devices based on this novel class of magnetic materials.

In conventional magnets, the resonant excitation of coherent phonons utilizing terahertz (THz) laser pulses has emerged as a powerful route for ultrafast magnetic control, enabling the exploration of nonthermal magnetic switching and hidden metastable states~\cite{Disa2021,Afanasiev2021,Disa2023,Ilyas2024,Zhang2024}.
The underlying mechanism is known as magnetophononics, where lattice vibrations couple to the spin degree of freedom primarily through the modulation of microscopic magnetic interactions, such as exchange coupling and magnetic anisotropy~\cite{Fechner2018,Gu2018,Afanasiev2021}.
While this mechanism is effective for interpreting targeted experimental fingerprints, it generally requires mode-resolved spin-phonon coupling strength as \textit{a priori} to predictively resolve the nonequilibrium magnetic response.
Such a parameter-dependent approach endows an opportunity to be reformulated when applied to altermagnets, where the magnetic state is fundamentally protected by the global crystal symmetry.
In this regard, theoretical studies on how the coherent phonon excitation interact with the altermagnetic order, particularly coupling to the nontrivial spin splitting, are urgently needed.

\begin{figure}[!ht]
    \centering
    \includegraphics[width=\linewidth]{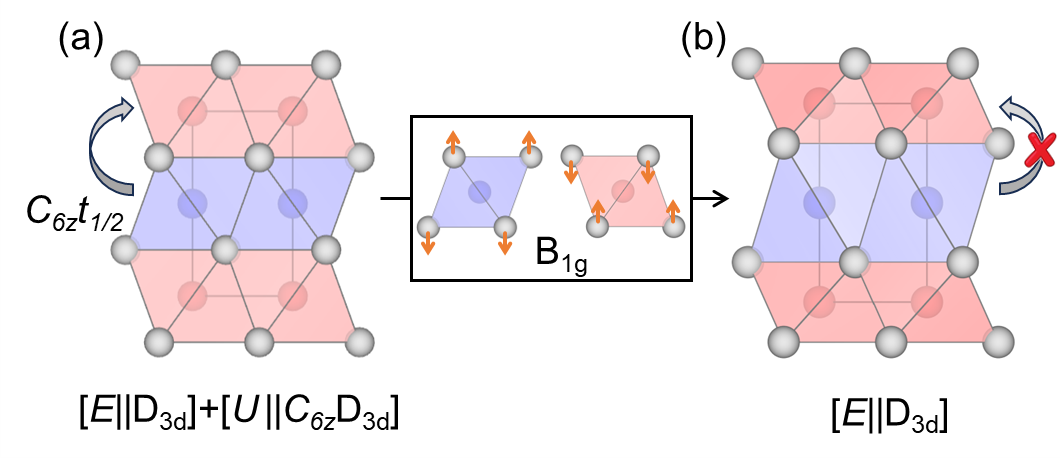}
    \caption{Schematic illustration of phonon-induced symmetry breaking in $\alpha$-MnTe. 
    (a) Crystal structure and spin group symmetry of $\alpha$-MnTe in its equilibrium altermagnetic phase. The opposite-spin Mn sublattices (blue and red) are connected by a non-symmorphic sixfold screw-axis rotation $C_{6z}t_{1/2}$. 
    The alternating spin splitting is described by the spin group symmetry $[E||D_{3d}]+[U||C_{6z}D_{3d}]$. 
    (b) Phonon-induced symmetry breaking upon excitation of a $B_{1g}$ mode with its vibrational eigenvector shown in the inset, which selectively annihilates the $C_{6z}$-related operations and reduces the spin group to $[E||D_{3d}]$. }
    \label{fig:schematic}
\end{figure}

\begin{figure*}[!ht]
    \centering
    \includegraphics[width=0.8\linewidth]{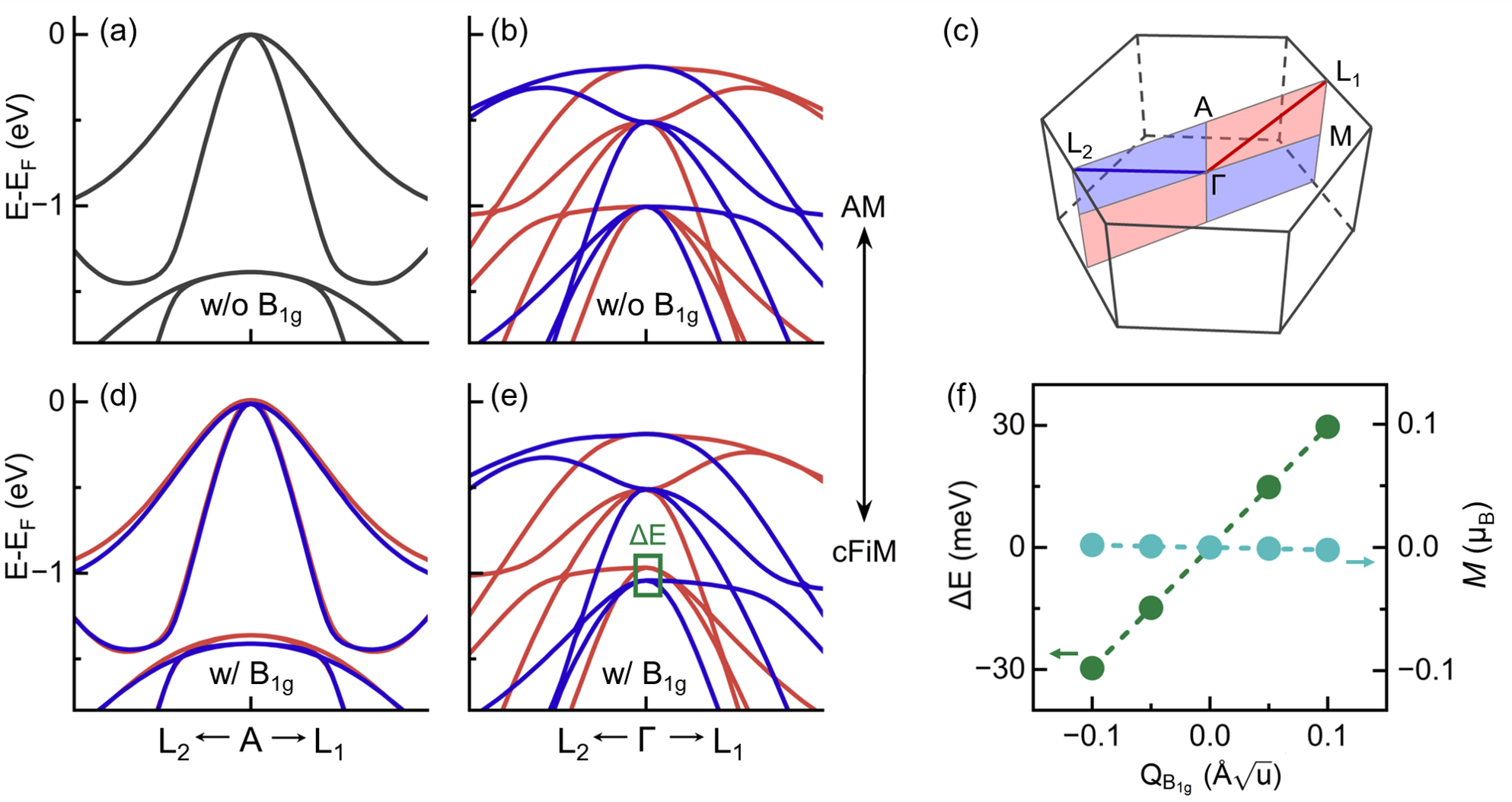}
    \caption{Symmetry-dictated transition from altermagnetism to compensated ferrimagnetism. 
    (a) and (b) Band structures of altermagnetic $\alpha$-MnTe. The spins are split within the normal plane of the Brillouin zone in (c) with the red and blue colours highlighting the alternating symmetry of the spin polarization, while remain degenerate along the high-symmetry paths (gray lines in (c)).
    (d) and (e) Renormalized band structure by the symmetry-breaking B$_{1g}$ phonon, where the spin degeneracies are lifted across the entire momentum space.
    In particular, the zone-center spin splitting $\Delta E$, which is region of our interest, is denoted by the green rectangle. 
    (f) Resultant $\Delta E$ and net magnetization ($M$) as a function of $B_{1g}$ phonon amplitude, implying the emergence of compensated ferrimagnetism characterized by a complete spin polarization without net magnetization.
    }
    \label{fig:altermag-ph}
\end{figure*}

In this Letter, we demonstrate the ultrafast switching of altermagnetism via altermagnetophononics, where coherent phonons act as symmetry-selective knobs for manipulating the altermagnetic state. Using $\alpha$-MnTe as a prototypical example, we show that specific phonon modes can lift the symmetry constraints protecting altermagnetism and introduce transitions to distinct magnetic phases.
Based on symmetry analysis, we find that B$_{1g}$-phonon-dressed MnTe exhibits a global spin splitting, even at the Brillouin zone center, yet maintains a vanishing net magnetization, as a direct manifestation of transition into a compensated ferrimagnetic state.
By selectively driving the phonon coherence along B$_{1g}$ mode via terahertz sum-frequency excitation, we realize a dynamical switching between different magnetic phases on a subpicosecond timescale.
Further, we demonstrate the generality of such symmetry-breaking mechanism, where the phase transition can also be induced by a simultaneous excitation of the infrared-active A$_{2u}$ and E$_{1u}$ phonons.
Finally, we extend the framework to a broader range of altermagnetic materials and demonstrate strategy for realizing reversible ferrimagnetic configuration in metallic altermagnet CrSb.
Our work complement the ultrafast control of altermagnet via coherent-phonon-induced selective symmetry breaking, which we hope can spark future experimental investigations.

\textit{Symmetry-dictated control scheme}--The defining feature of altermagnet, i.e., momentum-dependent nonrelativistic spin splitting, is governed by the spin-group symmetries  $[U||R]$, where the decoupled operations are \textit{U}, the spin-space rotation that reverses the spins, and \textit{R}, the real-space transformations that map the opposite-spin sublattices onto each other~\cite{AM202209,AM202212,Jiang2024,Chen2024SG,Xiao2024}.
Under equilibrium, these transformations ensure an alternating electronic structure satisfying $E_{\uparrow}(\bm{k})=E_{\downarrow}(R\bm{k})$, and the spins are degenerate merely at wavevectors where the requirement \textit{R\textbf{k}}=\textit{\textbf{k}} is fulfilled.
Such a symmetry-dictated spin texture implies that it is inherently susceptible to lattice perturbations that can explicitly alter the spatial operations \textit{R}.
In this regard, phonons, when pumped coherently to a finite amplitude, can act as a direct driving that selectively lowers the crystal symmetry to a subgroup compatible with their irreducible representations~\cite{Juraschek2017,Sie2019}.
The coherent-phonon-dressed nonequilibirum atomic geometry further leads to a reduced spin group, and correspondingly modifies the altermagnetism in a way that either preserves or lifts the spin degeneracy.
In particular, the system may settle into a different magnetic phase once the operations protecting the altermagnetic spin splitting, namely \textit{R}, are broken~\cite{Yuan2024}.

\begin{figure*}[!ht]
    \centering
    \includegraphics[width=0.8\linewidth]{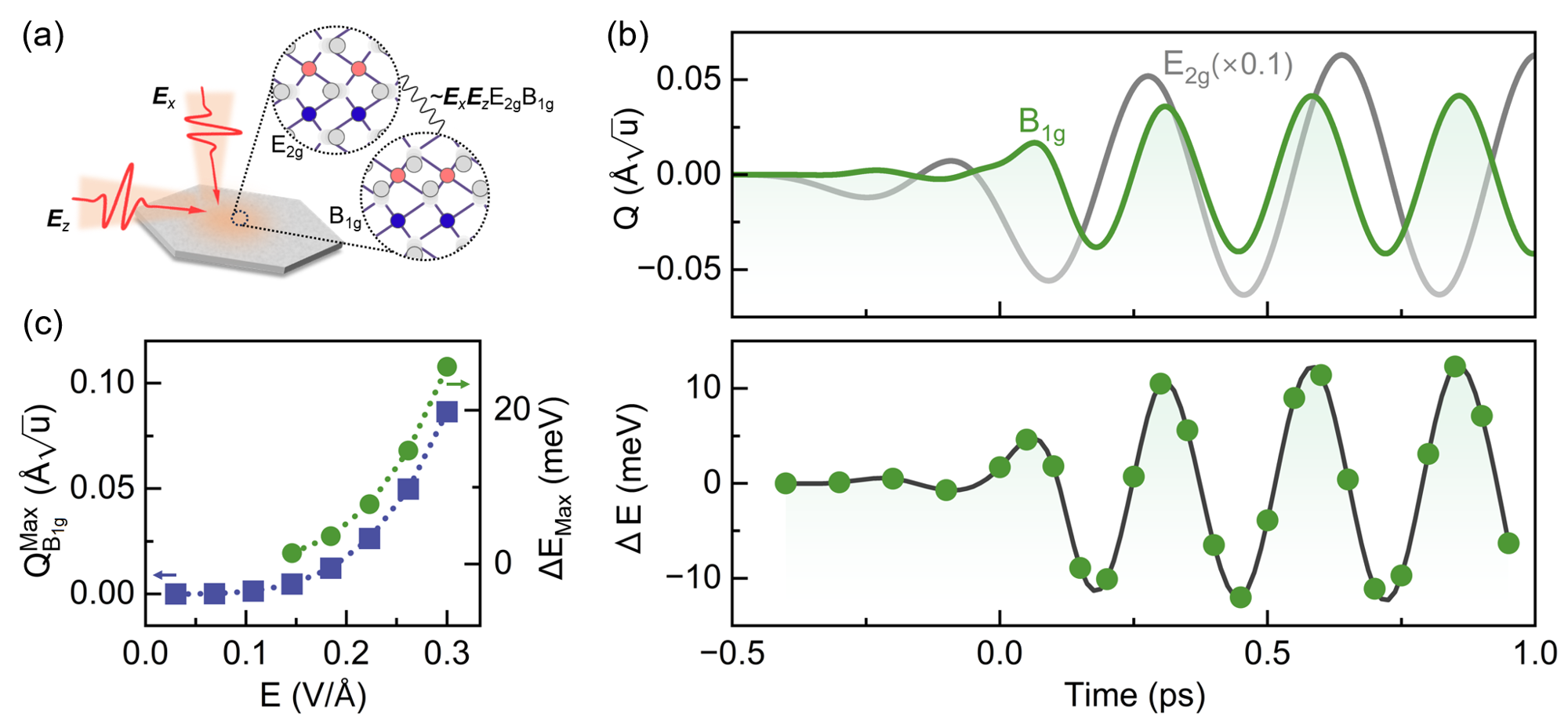}
    \caption{Ultrafast switching of altermagnetism via coherent phonon excitation. (a) Schematic illustration of the two-color terahertz sum-frequency excitation of the optical silent B$_{1g}$ mode mediated by a Raman-active E$_{2g}$ phonon. The insets show their corresponding atomic displacements.
    (b) Top panel: coherent  dynamics of the B$_{1g}$ (green) and E$_{2g}$ (gray) mode in response to terahertz excitation.
    Bottom panel: transient energy splitting $\Delta E$ calculated at each instantaneous lattice configuration set by Q$_{B_{1g}}$ and Q$_{E_{2g}}$. 
    The color shading is a guide to the eye to show the exclusive reliance of the magnetic switching on the B$_{1g}$ phonon dynamics.
    (c) Dependence of the maximum induced B$_{1g}$ phonon amplitude $Q_{B_{1g}}^{\rm Max}$ and zone-center spin splitting $\Delta E_{\rm Max}$ on THz-field strength.}
    \label{fig:dynamics}
\end{figure*}

To demonstrate this symmetry-based principle for altermagnetophononics, 
we now take the prototype altermagnet $\alpha$-MnTe as a paradigmatic example, focusing on the coherent-phonon-driven magnetic phase transition.
The symmetry condition of $\alpha$-MnTe is described by the nontrivial spin group with the symmetry transformations denoted as $[E||D_{3d}]+[U||C_{6z}D_{3d}]$~\cite{AM202209,AM202212}.
Here, the crystallographic group $D_{3d}$ contains all the operations that interchange atoms of the same spin, and $C_{6z}$ is the sixfold rotation connecting two sublattices holding antiparallel magnetic moments [Fig.~\ref{fig:schematic}(a)]~\cite{Gonzalez2023,Lee2024,Krempask2024,Liu2024}.
As shown in Figs.~\ref{fig:altermag-ph}(a)-(b), the corresponding ground-state band structure exhibits a nonrelativistic spin splitting within the normal plane (colored area in Fig.~\ref{fig:altermag-ph}(c)) and remains spin degenerate along the high-symmetry nodal lines such as $\mathrm{L}_2$-A-$\mathrm{L}_1$ (see Supplemental Material (SM) Sec.~S2 for details).

Based on the irreducible representation of its zone-center phonons, we first identify that the symmetry constraints can be lifted by exciting a single mode with B$_{1g}$ symmetry.
The B$_{1g}$ phonon comprises out-of-phase motion of the nonmagnetic Te atoms along the $c$-axis~[Fig.~\ref{fig:schematic}].
Such a motion renders the two opposite-spin Mn sublattices crystallographically inequivalent, and thus is odd under the spatial operations in $C_{6z}D_{3d}$.
Consequently, the spin group for MnTe with lattice distortion along the normal coordinate of Q$_{B_{1g}}$ is reduced to $[E||D_{3d}]$ [Fig.~\ref{fig:schematic}(b)], in a form corresponding to conventional ferro- or ferrimagnetism [Fig.~\ref{fig:schematic}(b)].

To quantitatively corroborate this symmetry-dictated phase transition, we show the renormalized band structure via the phonon-induced symmetry breaking as in Figs.~\ref{fig:altermag-ph}(d)-(e).
Notably, a $Q_{B_{1g}}-$dependent energy splitting emerges globally between the spin-up and spin-down manifolds, fundamentally deviating from the spin-momentum locking pattern in equilibrium (see Fig.~S2).
In particular, we find that the spin degeneracy can also be gaped out at $\Gamma$.
Here, we focus on the zone-center valence band with the most pronounced energy splitting ($\Delta E$), which we anticipate is amenable to photoemission detections.
As is shown in Fig.~\ref{fig:altermag-ph}(f), $\Delta E$ scales almost linearly with increased phonon amplitude and reaches $\sim$30~meV at Q$_{B_{1g}}$=0.1~\AA$\sqrt{u}$.
This implies the breakdown of the altermagnetic phase where $E_{\uparrow}(\Gamma)=E_{\downarrow}(\Gamma)$ is strictly enforced and symmetry-protected~\cite{Yuan2024}.
Intriguingly, although the unlocked spin splitting throughout the Brillouin zone suggests the emergence of a net magnetization, we reveal that the system maintains a robust antiferromagnetic order against the lattice distortion, as the gapped electronic structure of MnTe ensures that the spin-up and spin-down channel electrons are canceled out in its fully occupied valence bands.
It is noted that such a novel phase, also referred to as compensated ferrimagnetism (cFiM), has been recognized as a new type of order beyond altermagnetism, yet discussions on how it can be achieved are still limited~\cite{Yuan2024,Kawamura2024cFiM}.
In this regard, our findings propose a powerful pathway for the generation of cFiM from altermagnetism via phonon-induced symmetry breaking.

\textit{Coherent switching of altermagnetics}$-$Now, we discuss the realization of such symmetry-dictated control strategy in nonequilibrium regime by exciting the B$_{1g}$ phonon coherently.
Despite the success in manipulating magnetic order via THz-laser-driven infrared phonons~\cite{Disa2021,Afanasiev2021,Disa2023,Ilyas2024}, the B$_{1g}$ mode is neither infrared (IR) nor Raman active in the $D_{6h}$ point group of $\alpha$-MnTe, rendering it silent to standard IR absorption or Raman scattering techniques.
Moreover, a direct third-order coupling between the IR modes and the $B_{1g}$ mode is also symmetry forbidden, precluding the conventional route of nonlinear phononics based on cubic lattice anharmonicity (see SM Sec.~S3 for detailed analysis).

To overcome this restriction, we propose an alternative pathway via a two-color sum-frequency excitation mediated by a Raman-active E$_{2g}$ mode governed by the coupling potential
\begin{equation}
\begin{split}
        V&=\frac{1}{2}\Omega^2_{B_{1g}}Q_{B_{1g}}^2+\frac{1}{2}\Omega^2_{E_{2g}}Q_{E_{2g}}^2\\
    &+\varepsilon_0RE_1^2(\omega_1,t)Q_{E_{2g}}+\gamma E_1(\omega_1,t)E_2(\omega_2,t)Q_{E_{2g}}Q_{B_{1g}},
    \label{eq:potential}
\end{split}
\end{equation}
where $\Omega$ denotes the phonon frequencies, $R$ is the Raman tensor and $\gamma$ gives the nonlinear coupling coefficient.
The two pump pulses $E_1(\omega_1, t),E_2(\omega_2, t)$ are polarized respectively along the [100] and [001] direction of the crystal to ensure that the coupling term is symmetry allowed.
The sum-frequency scheme implies that their centering frequencies must satisfy $\omega_1=\Omega_{E_{2g}}/2$ and $\omega_1+\omega_2=\Omega_{E_{2g}}+\Omega_{B_{1g}}$ in order to achieve a high pump efficiency~\cite{Maehrlein2017,Juraschek2018,Kusaba2024}.
Here, the pump pulses are centered at $\omega_1=1.37$~THz and $\omega_2=5.00$~THz, respectively.
Such a set-up can effectively suppress the indiscriminate excitation of other branches due to either frequency mismatch or symmetry constraints (see SM Sec.~S3 for details).
More importantly, while the E$_{2g}$ phonon is necessary for the nonlinear process, the mode itself preserves the essential real-space symmetry operations in $C_{6z}D_{3d}$ and its excitation alone does not induce a transition to the compensated ferrimagnetic phase (see Fig.~S2).

By numerically solving the equations of motion corresponding to Eq.~(\ref{eq:potential}), we demonstrate the coherent phonon excitation via the sum-frequency pathway in Fig.~\ref{fig:dynamics}(b). Notably, the corresponding evolution of the spin texture reveals that the global lifting of spin degeneracy remains uniquely tied to the dynamical symmetry breaking induced by the B$_{1g}$ phonon coherence [bottom panel of Fig.~\ref{fig:dynamics}(b)], thereby maintaining a high degree of mode selectivity.
This exclusive reliance of the energy splitting on the B$_{1g}$ phonon dynamics further enables the external control over the nonequilibrium magnetic phase.
As shown in Fig.~\ref{fig:dynamics}(c), we observe that the maximum achievable phonon amplitude Q$^{\rm Max}_{B_{1g}}$ increases significantly with the THz field strength, resulting in an enhanced spin polarization in the induced cFiM phase.
These results demonstrate a more robust and universal pathway of \emph{altermagnetophononics}, for which the manipulation of altermagnetism can be guided and understood through crystal symmetry, without necessitating the calculation of complex magnetic interactions.

\textit{Discussion}$-$While we have identified the B$_{1g}$ mode in $\alpha$-MnTe as the dominant drive for the AM-to-cFiM transition, we emphasize that the framework of altermagnetophononics is not restricted to the excitation of one or two specific modes. In a more general sense, a similar effect can, in principle, be achieved through a suitable combination of multiple phonon modes when their collective vibrational pattern breaks the relevant spin group symmetries $[U||R]$. 

\begin{figure}
    \centering
    \includegraphics[width=0.85\linewidth]{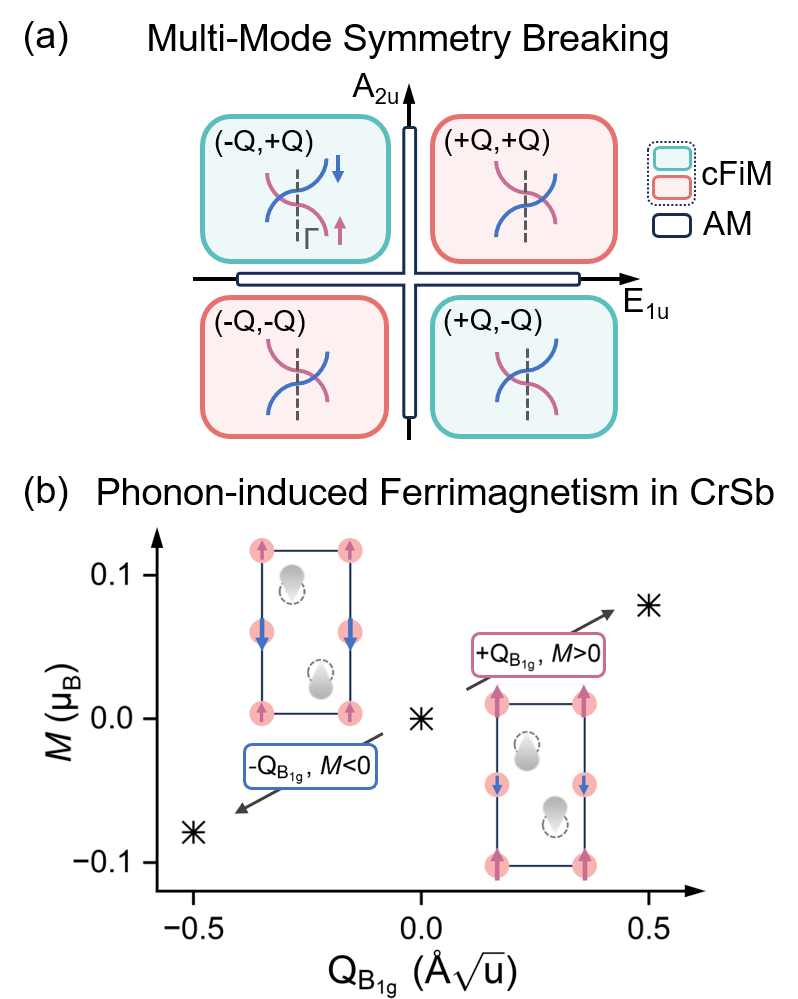}
    \caption{Generality of the symmetry-breaking mechanism. (a) Phase diagram in the phonon coordinate spanned by the infrared-active A$_{2u}$ and E$_{1u}$ modes in MnTe. The cFiM phase emerges only when the two modes are simultaneously excited, where the zone-center spin splitting can be effectively controlled by the relative phase of two modes. 
    (b) Altermagnetophononics in CrSb. A lattice distortion along the B$_{1g}$ phonon mode lifts the compensation between the antiferromagnetically aligned Cr moments, driving the system into a ferrimagnetic state displaying a finite net magnetization with its direction fully reversible by flipping the sign of effective displacement Q$_{B_{1g}}$.}
    \label{fig:generality}
\end{figure}

In the same case of $\alpha$-MnTe, we find that a simultaneous excitation of the two infrared-active A$_{2u}$ and E$_{1u}$ modes provides an alternative pathway to introduce the compensated ferrimagnetic phase $\alpha$-MnTe. 
While each mode individually preserves the symmetry elements protecting the altermagnetic state, their combined distortion effectively lowers the symmetry in a manner that allows a complete spin polarization (see Fig.~S2 for mode-dependent modulation in altermagnetism).
Moreover, the resulting zone-center energy splitting $\Delta E$ can become positive (negative) when the phonon amplitudes Q$_{A_{2u}}$ and Q$_{E_{1u}}$ are of the same (opposite) sign, as schematically illustrated in ~Fig.~\ref{fig:generality}(a).
This indicates that the relative phase of the two modes can offer an extra degree of freedom for multi-mode symmetry breaking, and this can be readily verified as the  infrared modes can be directly driven by a terahertz pulse.

Finally, we demonstrate the proposed framework can be broadly applied to identify phonon-induced magnetic phase transition across various altermagnetic materials.
In CrSb, a metallic system sharing the same crystal symmetry with MnTe ~\cite{Ding2024,Yang2025,Zhou2025}, we find that the excitation of B$_{1g}$ mode can also unlock the spin splitting throughout the Brillouin zone.
The symmetry-broken phase is characterized with a finite magnetization in contrary to that of MnTe.
Nevertheless, it is noteworthy that the induced net magnetic moment is tightly coupled to the structural degree of freedom, rendering its orientation fully reversible upon a sign change of the phonon displacement ( Fig.~\ref{fig:generality}(b), also see Sec.~S4 of the SM for more details).
Such distinct magnetic behavior can be attributed to the metallic electronic fillings of CrSb, where the spin-up and spin-down manifolds become asymmetric upon the B$_{1g}$ distortion.
These results suggest that, although the microscopic magnetic response may depend on the electronic structure of individual materials, the underlying mechanism based on symmetry-selective phonon driving remains robust.
Accordingly, similar phonon-induced magnetic phase transitions can also be achieved in altermagnets belonging to different symmetry classes, such as the RuO$_2$-type materials where the B$_{2g}$ mode plays the analogous role (see Fig.~S5 for application in RuO$_2$).

In conclusion, we have proposed altermagnetophononics, a symmetry-driven framework for ultrafast control of altermagnetic order via coherent phonon excitation.
We show that coherent phonons can break the real-space symmetry operations protecting the alternating spin texture, and drive the system into a distinct magnetic phases.
In $\alpha$-MnTe, excitation of either a single B$_{1g}$ mode or the infrared A$_{2u}$ and E$_{1u}$ modes simultaneously can induce a magnetic phase transition into a compensated ferrimagnetic order, featuring a complete spin polarization across the entire momentum space without net magnetization in the real space.
Further, by pumping the B$_{1g}$ phonon coherently via a terahertz sum-frequency route, we observe a highly coherent and mode-selective renormalization in the altermagnetic spin texture.
Such a symmetry-dictated mechanism can be further generalized taking into account of multi-mode symmetry breaking, and is applicable to various classes of altermagnetic materials, inducing ferrimagnetic order in metallic systems.
These findings shed light onto the magnetic response to coherent lattice dynamics in altermagnets, which we hope can advance both fundamental understanding and technological applications towards all-optical control of spintronic devices.

\soulregister{\ref}7
\begin{acknowledgments}
We acknowledge funding support from National Key Research and Development Program of China (Grant Nos.~2024YFA1408702, 2024YFF0508500, and 2021YFA1400200), National Natural Science Foundation of China (Nos.~12450401, 12474246), and Chinese Academy of Sciences (Nos. YSBR-047).
Y. W. acknowledges funding support by the Beijing Nova Program (No.~20250484898).
\end{acknowledgments}
\newpage
%

\end{document}